\documentclass{appolb}
\usepackage{graphicx}

\usepackage{dcolumn}
\usepackage{bm}
\usepackage{slashed}
\usepackage{amsmath,graphicx}
\usepackage[colorlinks=true,linktocpage=true,linkcolor=blue,citecolor=blue]{hyperref}
\usepackage{float}
\usepackage{nicefrac}
\usepackage[normalem]{ulem}
\usepackage{amsmath}
\usepackage{subfigure}

\newcommand{\beq}{\begin{eqnarray}}
\newcommand{\eeq}{\end{eqnarray}}

\newcommand{\bel}[1]{\begin{eqnarray}\label{#1}}
\newcommand{\eel}{\end{eqnarray}}

\newcommand{\rf}[1]{Eq.~(\ref{#1})}

\newcommand{\rfn}[1]{~(\ref{#1})}

\newcommand{\p}{\partial}

\newcommand{\tr}{\rm tr}

\newcommand{\f}[2]{\frac{#1}{#2}}
\newcommand{\onehalf}{{\nicefrac{1}{2}}}

\newcommand{\ed}{{\varepsilon}}       

\def\gmunu{g^{\mu\nu}}

\def\TmnU{T^{\mu \nu}}

\def\n0{n_{(0)}}
\def\e0{\varepsilon_{(0)}}
\def\P0{P_{(0)}}
\def\s0{s_{(0)}}

\def\fplusrsxp{f^+_{rs}(x,p)}

\def\bmu{\beta_\mu}

\def\umU{u^\mu}  
                         
\def\umL{u_\mu}

\def\unu{u^\nu}

\def\unuL{u_\nu}

\def\pmu{p^\mu}
\def\pnu{p^\nu}

\def\omnL{\omega_{\mu\nu}}
\def\omnU{\omega^{\mu\nu}}
\def\omnLbar{{\bar \omega}_{\mu\nu}}
\def\omnUbar{{\bar \omega}^{\mu\nu}}

\def\oabU{\omega^{\alpha\beta}}
\def\omnLD{{\tilde \omega}_{\mu\nu}}

\def\epsLmnbg{\epsilon_{\mu\nu\beta\gamma}}

\def\epsLmnab{\epsilon_{\mu\nu\alpha\beta}}
\def\epsUmnab{\epsilon^{\mu\nu\alpha\beta}}

\def\kmL{k_\mu}

\def\kmU{k^\mu}

\def\knL{k_\nu}

\def\omg{\omega^\gamma}

\def\omU{\omega^\mu}

\def\SmunuU{{\hat \Sigma}^{\mu\nu}}

\def\ubarrp{{\bar u}_r(p)}

\def\usp{u_s(p)}

\def\vbarsp{{\bar v}_s(p)}

\def\vrp{v_r(p)}

\def\g5{\gamma_5}

\def\fminusrsxp{f^-_{rs}(x,p)}

\def\slmnU{S^{\lambda, \mu \nu}}

\begin{document}
\title{Relativistic hydrodynamics of particles with spin $\onehalf\,\,\,$%
\thanks{Presented by WF at the ``Excited QCD 2017'' workshop, Sintra, Lisbon, Portugal, May 7-13, 2017.}%
}
\author{Wojciech Florkowski
\address{Institute of Nuclear Physics, PL-31342 Krakow, Poland \\
Jan Kochanowski University, PL-25406 Kielce, Poland \\
ExtreMe Matter Institute EMMI, GSI, D-64291 Darmstadt, Germany}
\\ \medskip
{Bengt Friman
}
\address{GSI Helmholtzzentrum f\"ur Schwerionenforschung, D-64291 Darmstadt, Germany}
\\ \medskip
{Amaresh Jaiswal
}
\address{GSI Helmholtzzentrum f\"ur Schwerionenforschung, D-64291 Darmstadt, Germany \\
School of Physical Sciences, National Institute of Science Education and Research, HBNI, Jatni-752050, India}
\\ \medskip
{Enrico Speranza
}
\address{GSI Helmholtzzentrum f\"ur Schwerionenforschung, D-64291 Darmstadt, Germany \\
Institut f{\"ur} Kernphysik, Technische Universit{\"a}t Darmstadt, D-64289 Darmstadt, Germany}
}
\maketitle
\begin{abstract}
A new hydrodynamic framework for particles with spin $\onehalf$,  based solely on the
conservation laws for charge, energy, momentum and angular momentum, is discussed. 
\end{abstract}
\PACS{25.75.-q, 24.10.Nz}
  
\section{Introduction}

In this  talk, we report on recent work~\cite{Florkowski:2017ruc}, where a novel hydrodynamic 
framework for particles with spin $\onehalf$ was introduced. The renewed interest in hydrodynamics 
of spinning particles is based on two facts: first, relativistic hydrodynamics forms the basic framework that is used to describe 
space-time evolution of matter created in relativistic heavy-ion collisions, studied experimentally at RHIC and 
the LHC~\cite{Florkowski:2017olj}, second, recently measurements of particle polarization in heavy-ion collisions have become
available~\cite{STAR:2017ckg}. Thus, it is tempting to combine these two topics  to explore polarization effects
in the context of hydrodynamic models (for a recent review of this and other related issues see, 
for example, Ref.~\cite{Wang:2017jpl} and references therein). 

\section{Local equilibrium distribution functions}

The main physics input for our approach is the definition of local equilibrium distribution functions for particles (plus signs) and antiparticles (minus signs) given in \cite{Becattini:2013fla}
\bel{fplusrsxp}
\fplusrsxp = \f{1}{2m} \ubarrp X^+ \usp,  \quad
\fminusrsxp = - \f{1}{2m} \vbarsp X^- \vrp.
\eel
Here $r,s = 1,2$ are spin indices, $u_r$ and $v_s$ are bispinors, and $X^{\pm}$ are the four by four matrices
\bel{XpmM}
X^{\pm} =  \exp\left[\pm \xi(x) - \bmu(x) \pmu \right] M^\pm,
\eel
where
\bel{Mpm}
M^\pm = \exp\left[ \pm \f{1}{2} \omnL(x)  \SmunuU \right].
\eel
Here we use the notation $\beta^\mu= \umU/T$ and $\xi = \mu/T$,  with the temperature $T$, chemical potential $\mu$, and the fluid four-velocity $\umU$ ($u \cdot u~=~1$). The quantity $\omnL$ is the polarization tensor, while  $\SmunuU$  is the spin operator expressed by the Dirac gamma matrices, $\SmunuU  = (i/4) [\gamma^\mu,\gamma^\nu]$. 

It is convenient to express the polarization tensor $\omnL$  in terms of the four-vectors $\kmU$ and $\omU$, 
\bel{omunuL}
\omnL \equiv \kmL \unuL - \knL \umL + \epsLmnbg u^\beta \omg . 
\eel
We can assume that both $k_\mu$ and $\omega_\mu$ are orthogonal to $\umU$ ($k \cdot u = \omega \cdot  u = 0$), hence 
\bel{kmuomu}
k_\mu = \omnL \unu, \quad \omega_\mu = \f{1}{2} \epsLmnab \, \omega^{\nu\alpha} u^\beta. 
\eel
We also define the dual polarization tensor
\bel{omunuLD}
\omnLD \equiv \f{1}{2} \epsLmnab  \oabU = \omega_\mu \unuL - \omega_\nu \umL +  \epsUmnab k_\alpha u_\beta.
\eel
It follows that  $\f{1}{2} \omnL \omnU = k \cdot k - \omega \cdot \omega$ and $\f{1}{2} \omnLD \omnU = 2 k \cdot \omega$. Using the constraint
\bel{conONE}
k \cdot \omega  = 0  
\eel
we find the compact form
\bel{Mpmexp}
M^\pm = \cosh(\zeta) \pm  \f{\sinh(\zeta)}{2\zeta}  \, \omnL \SmunuU  ,
\eel
where
\bel{zeta}
\zeta \equiv \f{1}{2} \sqrt{ k \cdot k - \omega \cdot \omega }.
\eel
%

\section{Basic physical observables}

The knowledge of the equilibrium distribution functions \rfn{fplusrsxp} allows us to compute the basic 
physical observables such as the charge and energy density, pressure, and entropy density. For the charge 
current we use the definition of Refs.~\cite{Becattini:2013fla, deGroot:1980} 
\bel{jmu}
N^\mu =  \int \f{d^3p}{2 (2\pi)^3 E_p}  \pmu \left[ \tr( X^+ ) - \tr ( X^- )  \right] = n \umU,
\eel
where ``$\tr$'' denotes the trace over spinor indices and $n$ is the charge density
\bel{nden}
n = 4 \, \cosh(\zeta) \sinh(\xi)\, \n0(T) = 2 \, \cosh(\zeta) \left(e^\xi - e^{-\xi} \right)\, \n0(T).
\eel
Here $\n0(T) = \langle(u\cdot p)\rangle_0$ is 
the number density of spinless, neutral Boltzmann particles, obtained using the thermal average
\bel{avdef}
\langle \cdots \rangle_0 \equiv \int \f{d^3p}{(2\pi)^3 E_p}  (\cdots) \,  e^{- \beta \cdot p} ,
\eel
where $p^0 = E_p = \sqrt{m^2 + {\bf p}^2}$ is the particle energy.

In the next step we calculate the energy-momentum tensor, again following Refs.~\cite{Becattini:2013fla,deGroot:1980} 
\bel{Tmn}
\TmnU =  \int \f{d^3p}{2 (2\pi)^3 E_p}  \pmu \pnu \left[ \tr( X^+ ) +  \tr ( X^- )  \right] = (\varepsilon + P ) \umU \unu - P \gmunu.
\eel
The energy density and pressure in \rfn{Tmn} are given by the formulas
\bel{enden}
\varepsilon = 4 \, \cosh(\zeta) \cosh(\xi) \, \e0(T) 
\eel
and
\bel{prs}
P = 4 \, \cosh(\zeta) \cosh(\xi) \, \P0(T), 
\eel
respectively. In analogy with the particle density $\n0(T)$, we define the auxiliary quantities 
$\e0(T) = \langle(u\cdot p)^2\rangle_0$ and $\P0(T) = -(1/3) \langle \left[ p\cdot p - 
(u\cdot p)^2 \right] \rangle_0$. We note that the energy-momentum tensor \rfn{Tmn} is symmetric and has the structure
characterizing perfect fluids. 

For the entropy current we use a straightforward generalization of the Boltzmann expression: 
\bel{s2}
S^\mu =  -\int \f{d^3p}{2 (2\pi)^3 E_p}  \, \pmu  \, \Big( \tr\left[ X^+ (\ln X^+ -1)\right]  +  \, \tr \left[ X^- (\ln X^- -1) \right] \Big) .
\eel
This leads to the entropy density which satisfies the equation
\bel{s}
s = u_\mu S^\mu = \f{\ed+P  - \mu \, n - \Omega w}{T} , 
\eel
where $\Omega = \zeta \, T$ and 
\bel{w}
w = 4 \, \sinh(\zeta) \cosh(\xi) \, \n0. 
\eel
The last equation suggests that $\Omega$ can be used as a thermodynamic variable of the  grand canonical potential, in addition to $T$  and $\mu$. Taking the pressure $P$ to be a function of $T, \mu$ and $\Omega$, $P=P(T,\mu,\Omega)$, one finds
\bel{dP}
s = \left.{\f{\p P}{\p T}}\right\vert_{\mu,\Omega}, \quad 
n = \left.{\f{\p P}{\p \mu}}\right\vert_{T,\Omega}, \quad 
w = \left.{\f{\p P}{\p \Omega}}\right\vert_{T,\mu}. 
\eel

\section{Hydrodynamic equations}

Hydrodynamic equations are first-order differential equations for the Lagrange multipliers
appearing in the local equilibrium distribution functions. Since we use the constraint \rfn{conONE}
and introduce $\Omega$ to parametrize the contraction $\omnL \omnU$, ten independent 
functions of space and time  are needed for a complete description.  These are chosen as:  
$T(x)$, $\mu(x)$, $\Omega(x)$, three independent components 
of $u^\mu(x)$, and the four remaining independent components of $\omnU(x)$. 

The conservation of energy and momentum  implies that
\bel{Tmncon1}
\p_\mu \TmnU = 0.
\eel
This equation can be split into  two parts, one longitudinal and the other transverse with
respect to $u^\mu$:
\bel{Tmncon2}
\p_\mu [(\ed + P) \umU ] &=& \umU \p_\mu P \equiv \f{dP}{d\tau},  \\
 (\ed + P ) \f{d \umU}{d\tau} &=& (g^{\mu \alpha} - u^\mu u^\alpha ) \p_\alpha P.
\eel
Evaluating the derivative on the left-hand side of the first equation one finds
\bel{snwcon}
T \,\p_\mu (s \umU) + \mu \,\p_\mu (n \umU) + \Omega \,\p_\mu (w \umU) = 0.
\eel
The term in the middle of the left-hand side vanishes due to charge conservation,
\bel{ncon}
\p_\mu (n \umU)=0.
\eel
Thus, in order to have conservation of entropy in our system, $\p_\mu (s \umU)~=~0$ (for the 
perfect-fluid description we are aiming at), we demand that
\bel{wcon}
\p_\mu (w \umU) = 0.
\eel
Equations \rfn{Tmncon1}, \rfn{ncon} and \rfn{wcon} form a closed system of six equations for six unknowns: $T(x)$, $\mu(x)$, $\Omega(x)$ and three components of $u^\mu(x)$. Since they do not determine the time evolution of the individual components
of the polarization tensor, we dub them the equations for the hydrodynamic background.

\section{Spin dynamics}

Our approach is based on the conservation of the angular momentum
in the form $\p_\lambda J^{\lambda, \mu\nu}=0$, where $J^{\lambda, \mu\nu} = L^{\lambda, \mu\nu} 
+ S^{\lambda, \mu\nu}$ with $L^{\lambda, \mu\nu}=x^\mu T^{\nu\lambda}- x^\nu T^{\mu\lambda}$
being the orbital angular momentum and $S^{\lambda, \mu\nu}$  the spin tensor. 
Since the energy-momentum tensor \rfn{Tmn} is symmetric,  the conservation law $\p_\lambda J^{\lambda, \mu\nu}=0$ 
implies conservation of the spin tensor $\slmnU$~\cite{Hehl:1976vr},
\bel{spincon1}
\p_\lambda \slmnU = 0.
\eel
For $\slmnU$ we use the form discussed in \cite{Becattini:2009wh}
\bel{st11}
\slmnU = \!\!\int\!\!\f{d^3p}{2 (2\pi)^3 E_p} \, p^\lambda \, {\tr} \left[(X^+\!-\!X^-) \SmunuU \right]  =  \frac{w u^\lambda}{4 \zeta}  \omega^{\mu \nu} .
\eel

Using the conservation law for the spin density and introducing the rescaled 
polarization tensor $\omnUbar = \omnU/(2\zeta)$, we find
\bel{st12}
u^\lambda \p_\lambda \, \omnUbar = \f{d\omnUbar }{d\tau} = 0.
\eel
Since, $\omnUbar$ is antisymmetric, \rf{st12} with the normalization condition 
\bel{norm}
\omnLbar \, \omnUbar = 2
\eel
yields five independent equations.  If the condition \rfn{conONE} is fulfilled on the initial
hypersurface, it  remains fulfilled at later times, provided  \rf{st12}  holds. Hence, \rf{st12} used with \rfn{conONE} and \rfn{norm} yields four additional equations that  are needed to determine the space-time evolution of a spinning fluid. 
In Ref.~\cite{Florkowski:2017ruc} we have shown that this framework has  a vortex-like solution that corresponds 
to global equilibrium studied in Refs.~\cite{Becattini:2013fla,Becattini:2009wh}.

\section{Closing remarks}

In this work we have described a new hydrodynamic approach to relativistic perfect-fluid hydrodynamics of 
particles with spin $\onehalf$. The system of hydrodynamic equations follows directly from the conservation
laws for charge, energy, momentum and angular momentum.  An important ingredient of our approach 
is the form of the spin tensor defined by \rf{st11} that allows for the construction of a consistent system
of equations. We note that the form \rfn{st11} differs from those
used in \cite{Becattini:2013fla} and \cite{deGroot:1980}, respectively.

\bigskip
{\bf Acknowledgments:} This research was supported in part by the ExtreMe Matter Institute 
EMMI at the GSI Helmholtzzentrum f\"ur Schwerionenforschung, 
Darmstadt, Germany.


\begin{thebibliography}{100}
\expandafter\ifx\csname url\endcsname\relax \def\url#1{{\tt #1}}\fi
\expandafter\ifx\csname urlprefix\endcsname\relax\def\urlprefix{URL}\fi
\providecommand{\eprint}[2][]{\url{#2}}

\bibitem{Florkowski:2017ruc} 
  W.~Florkowski, B.~Friman, A.~Jaiswal and E.~Speranza,
  arXiv:1705.00587 [nucl-th].
  
\bibitem{Florkowski:2017olj}
  W.~Florkowski, M.~P.~Heller and M.~Spalinski,
  arXiv:1707.02282 [hep-ph].
    
\bibitem{STAR:2017ckg}
  L.~Adamczyk {\it et al.} [STAR Collaboration],
  arXiv:1701.06657 [nucl-ex].  
  
  \bibitem{Wang:2017jpl}
  Q.~Wang,
  arXiv:1704.04022 [nucl-th].

\bibitem{Becattini:2013fla} 
  F.~Becattini, V.~Chandra, L.~Del Zanna and E.~Grossi,
  Annals Phys.\  {\bf 338} (2013) 32.
  
\bibitem{deGroot:1980}
S. R. de~Groot, W.A. van~Leeuwen, and Ch.G. van~Weert, 
Relativistic Kinetic Theory: Principles and Applications,
(North-Holland, Amsterdam, 1980).

\bibitem{Hehl:1976vr}
  F.~W.~Hehl,
  Rept.\ Math.\ Phys.\  {\bf 9} (1976) 55.

\bibitem{Becattini:2009wh}
  F.~Becattini and L.~Tinti,
  Annals Phys.\  {\bf 325} (2010) 1566.
    
\end{thebibliography}
\end{document}